\documentclass[a4paper,12pt]{article}
\usepackage{amsmath,amsthm,amssymb,bbm}

\usepackage[english]{babel}

\title{\bf Sub-shot-noise quantum metrology\\ with entangled identical particles}

\author{F. Benatti$^{a,b}$, 
R. Floreanini$^{b}$, U. Marzolino$^{a,b}$\\
\\
\small ${}^a$Dipartimento di Fisica, Universit\`a di Trieste, 
34014 Trieste, Italy\\
\small ${}^b$Istituto Nazionale di Fisica Nucleare, Sezione di Trieste,
34014 Trieste, Italy}

\date{\null}

\begin{document}

\maketitle

\begin{abstract}
\noindent
The usual notion of separability has to be reconsidered when
applied to states describing identical particles.
A definition of separability not related to any {\it a priori} Hilbert space
tensor product structure is needed: this can be given
in terms of commuting subalgebras of observables.
Accordingly, the results concerning the use of 
the quantum Fisher information in quantum metrology are
generalized and physically reinterpreted.
\end{abstract}

\section{Introduction}

Quantum interferometry is one of the most useful measurement technique used in quantum metrology:
it exploits the high sensitivity of certain quantum systems to small changes of external conditions,
due to their intrinsic quantum coherence. Although various implementations of this general
measurement procedure have been realized, the aim is always that of estimating the relative phase of
the two ``arms'' of the interferometer and great efforts have been pursued to enhance the sensitivity
of this phase determination \cite{Knight,Cronin}.

The accuracy of the whole estimation procedure \cite{Helstrom}-\cite{Paris}
depends on the number of resources at disposal, 
{\it i.e.} on the total number $N$ of particles involved in the interferometric measure.
It turns out that the minimum uncertainty in the phase estimation that can be achieved by feeding the
interferometer with ``classical'' states can not exceed the shot noise limit, also called
the standard quantum limit, where the precision scales as $1/\sqrt{N}$.

Various strategies have been devised in order to beat this limit, typically adopting non-standard
measurement protocols that make use of quantum correlated (entangled), non-classical states.%
\footnote{The literature on the subject is vast, dealing both with optical
as well as atom interferometers; for a partial list, 
see \cite{Caves1}-\cite{Oberthaler} and references therein.}
In this way, at least in line of principle, the so called Heisenberg limit in accuracy can be reached,
where the uncertainty in the phase estimation scales as $1/N$ rather than $1/\sqrt{N}$.

A particularly promising implementation of an interferometer that could reach such a high
sensitivity is that of a Bose-Einstein condensate (BEC) \cite{Stringari}-\cite{Leggett}
trapped in a double-well optical potential.
Although similar to more traditional constructions that use beams of particles
travelling along separate paths, this new realization of the interferometry paradigm
allows a precise control on the preparation and dynamics of the system.

The clear advantage in using BEC-based interferometers is that all particles in the apparatus,
being in a condensed phase, share the same quantum state; this fact, together with the possibility
of preparing them in a suitably entangled state, have been shown to lead to the possibility 
of phase estimation with sub-shot-noise accuracy \cite{Bouyer}-\cite{Briegel}. 
However, these particles, being spatially confined bosons, behave as identical particles,
a fact that seems to have not been properly taken into account 
in the derivation of the above mentioned results.

Purpose of the present investigation is to study the effects of particle indistinguishability
in relation to the metrological use of BEC-based interferometers for phase estimation.
Since entanglement appears to be a fundamental resource in reaching sub-shot-noise accuracies,
one first needs to generalize this notion to states describing identical particles.
This is presented in Section 2; there, we shall see that one needs to abandon the
usual definition of separability expressed in terms of tensor product of one-particle
Hilbert spaces, in favor of a more general definition based on the choice of commuting
algebras of observables. In Section~3 we shall explicitly discuss how this generalized
notion of entanglement, based on a dual description in terms of operators rather than states,
works in the case of a BEC confined in a double-well trap. Finally, in the light of this
discussion, in Section 4 we shall critically re-examine the theoretical results
concerning the use of the notion of quantum Fisher information
in getting sub-shot-noise accuracies in quantum metrological phase estimation.
The final Section 5 contains further discussions and remarks.

\section{Identical particles entanglement}

The usual notion of entanglement for states of a system of $N$ distinguishable particles makes use
of the natural tensor product structure of the $N$-body system. The total Hilbert space
$\cal H$ of the system can in fact be decomposed as 
${\cal H}={\cal H}_1\otimes{\cal H}_2\otimes\ldots\otimes{\cal H}_N$ in terms of the various
single-particle Hilbert spaces ${\cal H}_i$, $i=1, 2\ldots, N$. Then a state for the
$N$-body system, represented by a density matrix $\rho$ acting on $\cal H$, is said to
be separable if it can be written as a convex combination of single-particle states,
\begin{equation}
\rho=\sum_k p_k\, \rho_k^{(1)}\otimes\rho_k^{(2)}\otimes\ldots\otimes\rho_k^{(N)}\ ,\qquad p_k\geq 0\ ,\quad
\sum_k p_k=1\ ,
\label{1}
\end{equation}
where $\rho^{(i)}$ represents a density matrix for the $i$-th particle; otherwise, the state
$\rho$ for the $N$ distinguishable particles is said to be entangled.

The tensor product structure used in (\ref{1}) reflects on one hand the natural multi-partition
of the system into its $N$ parts, and on the other the possibility of identifying linear local operations
on the system, {\it i. e.} maps of the form%
\footnote{In order to be physically meaningful, these linear transformations must be realized by
trace-preserving, completely positive maps \cite{Alicki,Takesaki}; 
only in this case, normalization, ${\rm Tr}[\rho]=1$,
and positivity, $\rho\geq0$, of the states $\rho$ of the $N$-body system are assured in any situation 
(for a discussion, see \cite{Benatti}).}
\begin{equation}
\Lambda=\Lambda^{(1)}\otimes\Lambda^{(2)}\otimes\ldots\otimes\Lambda^{(N)}\ ,
\label{2}
\end{equation}
where the linear map $\Lambda^{(i)}: \rho^{(i)}\mapsto \Lambda^{(i)}\big[\rho^{(i)}\big]$ 
affects only the $i$-th particle density matrix.
 
Coming now to the case of a system of $N$ identical particles, one easily realizes
that states of the form (\ref{1}) are not allowed quantum states for the system.
Indeed, if one could assign a density matrix $\rho^{(1)}$ to the first particle,
$\rho^{(2)}$ to the second one and so on, we would have a way to distinguish them
by means of their identifying states. 

On the other hand, recalling the standard
rules of quantum mechanics, a pure state $|\psi\rangle$ of $N$ identical particles
must be a symmetric or antisymmetric combination of tensor products of $N$-single
particle vector states, while a mixed state, {\it i.e.} a density matrix, must be
a linear convex combination of projections $|\psi\rangle\langle\psi|$ onto such
symmetrized or antisymmetrized vectors~\cite{Feynman,Sakurai}.%
\footnote{Similarly, also the observables must be symmetric or antisymmetric
under the permutation of all particles; in particular, collective,
$N$-body operators are surely admissible observables (see also the note on page 9).}

In particular, this result excludes the possibility of curing 
the intrinsic distinguishability of the
particles when described by states as in
(\ref{1}), by symmetrizing or antisymmetrizing them:
indeed, in the end, this would always be a mixture of states where
particles can be effectively distinguished. In other words, density matrices describing
identical particles must be made of projections onto pure states
of the appropriate (symmetrized or antisymmetrized) Hilbert space.
As a consequence, not even the so-called ``symmetric states'' of the form
\begin{equation}
\rho=\sum_k p_k\, \rho_k\otimes\rho_k\otimes\ldots\otimes\rho_k\ ,
\label{3}
\end{equation}
where, unlike in (\ref{1}), all particles in each term of the convex combination
are described by the same one-particle density matrix, are in general 
admissible states for a system of identical particles.

The example of a system of two qubits can further clarify these points.
The Hilbert space of two distinguishable qubits is $\bold{C}^2\otimes\bold{C}^2\equiv\bold{C}^4$,
and therefore four-dimensional. Instead, the Hilbert space for two identical
qubits is a symmetric three-dimensional subspace of $\bold{C}^4$ in the case of bosons,
or an antisymmetric one-dimensional subspace for fermions. Given an orthonormal
basis in $\bold{C}^4$, $|+,+\rangle$, $|+,-\rangle$, $|-,+\rangle$, $|-,-\rangle$,
where $+$ and $-$ indicates the two single-qubit levels,
the symmetric sector is spanned by $|+,+\rangle$, $|+,+\rangle$, 
and the combination $(|+,-\rangle + |-,+\rangle)/\sqrt{2}$, while the antisymmetric one
is generated by $(|+,-\rangle - |-,+\rangle)/\sqrt{2}$. It is now easy to realize that 
density matrices of the type (\ref{1}), 
$\rho=\sum_k p_k\, \rho_k^{(1)}\otimes\rho_k^{(2)}$, can not describe
states of two identical qubits. Indeed, a qubit state $\rho_k^{(i)}$ is represented by a $2\times2$ density
matrix, so that any element $\rho_k^{(1)}\otimes\rho_k^{(2)}$ in the previous convex sum
is a $4\times4$ matrix, as is the resulting combination $\rho$. 
As such, it can not be written in general as a convex combination of solely 
projections onto symmetric states or in terms of the projection 
onto the antisymmetric one, as it would be instead required for density matrices
describing two identical qubits.
Clearly, the same conclusion applies also for two qubit states of the form
$\sum_k p_k\, \rho_k\otimes\rho_k$.

Given that the notion of separability based on the decomposition (\ref{1}) is no longer
applicable in the case of a system of identical particles, one is forced to generalize
it. This problem has already been addressed and discussed in the literature, although
in different physical contexts, {\it e.g.} see \cite{Ghirardi}-\cite{Buchleitner}. 
The way out seems to reside in a ``dual'' point of view,
where emphasis is given to the algebra of observables, instead of the set of states;
the connection between these two ``points of view'' is given by the expectation value map
that allows to express the average of an observable ${\cal O}$ as its trace with the
corresponding density matrix, $\langle{\cal O}\rangle={\rm Tr}[\rho\,  {\cal O}]$.
We shall first give an abstract definition of this generalized notion of separability
and then, in the next Section, apply it to the specific case of condensed 
bosons trapped in a double-well potential.

Let us consider a system of particles
whose pure states span the Hilbert space ${\cal H}$ and denote by
${\cal B}({\cal H})$ the algebra of all bounded operators on it; the observables 
of the system clearly belong to this algebra. We shall introduce the notion of 
(bipartite) separability by considering couple of commuting subalgebras of ${\cal B}({\cal H})$
instead of focusing on partitions of the system quantum states. We then introduce the
following definitions:

\begin{itemize}
\item{} An {\sl algebraic bipartition} of the algebra ${\cal B}({\cal H})$ is any pair
$({\cal A}_1, {\cal A}_2)$ of {\sl commuting subalgebras} of ${\cal B}({\cal H})$.%
\footnote{Notice that the two subalgebras ${\cal A}_1$ and ${\cal A}_2$ need not reproduce the whole
algebra ${\cal B}({\cal H})$, {\it i.e.} in general 
\hbox{${\cal A}_1 \cup {\cal A}_2\subset {\cal B}({\cal H})$}.
In this respect, the term ``bipartition'' is not strictly appropriate and has been adopted
for sake of simplicity. However, in the case of the system discussed below, the considered mode
partitions actually generate the whole algebra ${\cal B}({\cal H})$.}
\item{} An element (operator) of ${\cal B}({\cal H})$ is said to be {\sl local} with respect to
the bipartition $({\cal A}_1, {\cal A}_2)$ if it is the product $A_1 A_2$ of an element 
$A_1$ of ${\cal A}_1$ and another $A_2$ of ${\cal A}_2$.
\item{} A state $\omega$ on the algebra ${\cal B}({\cal H})$ will be called {\sl separable} with
respect to the bipartition $({\cal A}_1, {\cal A}_2)$ if the expectation $\omega(A_1 A_2)$ 
of any local operator $A_1 A_2$ can be decomposed into a linear convex combination of
products of expectations:
\begin{equation}
\omega(A_1 A_2)=\sum_k\lambda_k\, \omega_k^{(1)}(A_1)\, \omega_k^{(2)}(A_2)\ ,\qquad
\lambda_k\geq0\ ,\quad \sum_k\lambda_k=1\ ,
\label{4}
\end{equation}
where $\omega_k^{(1)}$ and $\omega_k^{(2)}$ are states on ${\cal B}({\cal H})$;
otherwise the state $\omega$ is said to be {\sl entangled} with respect the bipartition
$({\cal A}_1, {\cal A}_2)$. 
\end{itemize}

As we shall see in the following, this generalized definition of separability 
results meaningful both for systems composed
of distinguishable particles and for those made of identical ones. In particular, in the case of two 
finite-level systems, as the two qubits example discussed before, the above definition reduces
to the usual notion of separability. Indeed, it is enough to choose the subalgebras
${\cal A}_1$ and ${\cal A}_2$ to coincide with the $2\times 2$ matrix algebras
of the single-qubits and take as operation of expectation the usual trace operator
over the corresponding density matrix,
$\omega(A_1 A_2)={\rm Tr}[\rho\, A_1\otimes A_2]$; clearly, this expectation can be written
as in (\ref{4}) only if the system state $\rho$ is as in (\ref{1}) with $N=2$. 
Finally, generalization
of the above definition to the case of more than two partitions is straightforward.

\section{Identical bosons in a double-well trap}

We shall henceforth focus on the study of a system of ultracold bosonic atoms trapped
in an optical lattice. As mentioned in the Introduction, the high degree of accuracy and
sophistication reached in manipulating such systems allow very accurate interferometric
measurements, making them a very attractive set-up for exploring and testing
new paradigms in quantum metrology and many-body physics \cite{Lewenstein2,Bloch}.

The dynamics of cold atoms confined in an optical trap can be very well approximated
by a Bose-Hubbard type Hamiltonian \cite{Milburn,Jaksch}; 
in the case of a double-well potential, it reads:
\begin{equation}
H_{BH}=\varepsilon_1\, a^\dagger_1a_1\,+\,\varepsilon_2\, a^\dagger_2a_2\,+\,
U\,\Bigl[(a^\dagger_1a_1)^2+(a^\dagger_2a_2)^2\Bigr]\,-\,J\,(a^\dagger_1a_2+a_1a^\dagger_2)\ ,
\label{5}
\end{equation}
where $a_{1,2}$, $a_{1,2}^\dag$ annihilate and create atom states in the first, 
second well, respectively, and satisfy the 
Bose commutation relations $[a_i\,,\,a^\dag_j]=\delta_{ij}$, $i,j=1,2$.
Of the contributions to $H_{BH}$, the last one corresponds to a
hopping term depending on the tunneling amplitude $J$, the first two
are due to the trapping potential and are proportional to the depth $\varepsilon_{1,2}$ of
the wells. Finally,
the third term, quadratic in the number operators $a^\dag_i
a_i$, takes into account repulsive Coulomb interactions inside each well.

We are using here a formalism of second-quantization, where
the creation operator $\hat{\psi}^\dag(x)$ of an atom at position $x$ 
can be decomposed in general as
\begin{equation}
\hat{\psi}^\dag(x)=\sum_{i=1}^\infty\phi_i^*(x)\,a^\dag_i\ ,
\label{6}
\end{equation}
and $a^\dag_i$ creates an atom in the state $|\phi_i\rangle=a^\dag_i |0\rangle$ with wavefunction
$\phi_{i}(x)=\langle x | \phi_i\rangle$; the set $\{|\phi_i\rangle\}_{i=1}^\infty$ is therefore 
a complete set of orthonormal single-particle atom states.
The Bose-Hubbard Hamiltonian~(\ref{5}) results from a
\textit{tight binding approximation}, where only the first two of the
basis vector are relevant; in this case
$\phi_{1,2}(x)$ are orthogonal functions, $\phi_1$ localized
within the first well, $\phi_2$ within the second one.

The total number $N$ of atoms is conserved by (\ref{5}).
Therefore, the Hilbert space of the system is $N+1$-dimensional and can be
spanned by Fock states of the form
\begin{equation}
|{k,N-k}\rangle:=\frac{(a_1^\dag)^k(a_2^\dag)^{N-k}}{\sqrt{k!}\sqrt{(N-k)!}}|0\rangle\ ,
\qquad 0\leq k\leq N\ ,
\label{7}
\end{equation}
describing the situation in which the first well is filled with $k$ atoms,
while the other one contains $N-k$ particles. They are obtained by the action of the
creation operators on the vacuum. These states turn out to be eigenstates
of the Bose-Hubbard Hamiltonian when the tunneling term can be neglected.

Notice that in this formalism, symmetrization of the elements of the Hilbert space,
as required by the identity of the particles filling the two wells,
is automatically guaranteed by the commutativity of the two creation operators \cite{Feynman}.
Furthermore, all polynomials in $a_1$, $a_1^\dag$
and similarly all polynomials in $a_2$, $a_2^\dag$ (together with their respective norm closures), 
form two algebras,
${\cal A}_1$, respectively ${\cal A}_2$, that result commuting: $[{\cal A}_1, \, {\cal A}_2]=\,0$;
they are subalgebras of the algebra $\cal A$ of all operators on the Fock space spanned by 
the states (\ref{7}).%
\footnote{Strictly speaking, polynomials in bosonic creation and annihilation operators are
not bounded, and the so-called Weyl algebras of the corresponding exponentials should be used;
however, the following discussion is not affected by working with algebras of unbounded operators.}
According to the definition introduced at the end of the
previous Section, they define a bipartition of $\cal A$ and therefore 
can be used to provide the notion of separability for the states
describing the identical atoms in the trap, 
generalizing the one usually adopted for distinguishable particles.

With respect to this natural mode bipartition, $({\cal A}_1, {\cal A}_2)$, the Fock states
(\ref{7}) turn out to be separable states. Indeed, for any polynomial operator $A_1(a_1,a_1^\dag)\in {\cal A}_1$
and $A_2(a_2,a_2^\dag)\in {\cal A}_2$, the expectation value of the product $A_1 A_2$
on such states can always be written as the product of the separate averages of $A_1$ and of $A_2$;
explicitly, one has:
\begin{eqnarray}
\nonumber
\langle {k,N-k}|A_1 A_2|{k,N-k}\rangle &=& {1\over k! (N-k)!}
\langle 0| a_1^k\, A_1\, (a_1^\dag)^k |0\rangle\
\langle 0| a_2^{N-k}\, A_2\, (a_2^\dag)^{N-k} |0\rangle\\
&=&\langle k| A_1 |k\rangle\ \langle N-k| A_2 |N-k\rangle\ ,
\label{8}
\end{eqnarray}
where $|k\rangle:=(a_1^\dag)^k/\sqrt{k!} |0\rangle$ and 
$|N-k\rangle:=(a_2^\dag)^{N-k}/ \sqrt{(N-k)!}|0\rangle$ 
are single-mode Fock states.

As a consequence, separable with respect to the bipartition $({\cal A}_1, {\cal A}_2)$ are also those
mixed states that are diagonal with respect to the Fock basis (\ref{7}),
{\it i.e.} density matrices of the form:
\begin{equation}
\rho=\sum_{k=0}^N p_k\, |k,N-k\rangle\langle k, N-k |\ ,\qquad
p_k\geq 0\ ,\quad \sum_{k=0}^N p_k=1\ .
\label{9}
\end{equation}

Actually, all states separable with respect to the bipartition $({\cal A}_1, {\cal A}_2)$ must
be in diagonal form with respect to the Fock basis. An abstract proof of this fact can be
easily given. In fact, one first observes that the algebra generated
by the polynomials in the operators $a_1$, $a_1^\dag$ and $a_2$, $a_2^\dag$ coincides
with the whole algebra of operators $\cal A$. Then, being pure projections, the states
$|k,N-k\rangle\langle k, N-k |$ are the only extremal ones in the convex set of
states satisfying the separability condition (\ref{4}); thus, all others are necessarily
a convex combination of them. 

Nevertheless, a more direct proof can also be explicitly worked out.
A generic density matrix representing a state for a two-mode system of $N$ identical bosons
is of the form:
\begin{equation}
\rho=\sum_{k,l=0}^N \rho_{kl}\, |k,N-k\rangle\langle l, N-l |\ ,\qquad
\rho_{kk}\geq 0\ ,\quad \sum_{k=0}^N \rho_{kk}=1\ .
\label{10}
\end{equation}
Assume now the state $\rho$ to be separable with respect to the bipartition $({\cal A}_1, {\cal A}_2)$;
this means that for any local operator $A_1 A_2$, with $A_1\in{\cal A}_1$ and $A_2\in{\cal A}_2$,
one must have:
\begin{equation}
{\rm Tr}[\rho\, A_1 A_2]=\sum_m \lambda_m {\rm Tr}[\rho^{(1)}_m A_1]\, {\rm Tr}[\rho^{(2)}_m A_2]\ ,
\label{11}
\end{equation}
where $\rho^{(i)}_m$, $i=1,2$, are density matrices of the generic form (\ref{10}).
Choose now $A_1=(a_1^\dag)^m\, a_1^n$, with $m<n$ and $A_2=(a_2^\dag)^r\, a_2^s$,
with $s<r$ and $m+r=n+s$; then, each factor in the sum above
identically vanish. Indeed, one finds:
\begin{equation}
{\rm Tr}[\rho_m^{(1)}\, A_1]=
\sum_{k,l=0}^N \rho_{m,kl}^{(1)}\, \langle k,N-k| A_1 |l,N-l\rangle=
\sum_{k=0}^N \rho_{m,kk}^{(1)}\, \langle k| A_1 |k\rangle\equiv\,0\ ,
\label{12}
\end{equation}
and similarly, ${\rm Tr}[\rho^{(2)}_m A_2]\equiv\,0$. However, using (\ref{10}),
one also have:
\begin{equation}
{\rm Tr}[\rho\, A_1 A_2]=\sum_{k,l=0}^N \rho_{kl}\, \langle k| A_1 |l\rangle\,
\langle N-k | A_2 | N-l\rangle=\sum_{k=n}^{N-s}\rho_{k,k-n+m}\ \alpha_{m,n}(k)\, \beta_{r,s}(k)\ ,
\label{13}
\end{equation}
where the coefficients $\alpha$ and $\beta$ are explicitly given by:
\begin{eqnarray}
\nonumber
&& \alpha_{m,n}(k)=\Big[k(k-1)\cdots(k-n+1)\cdot(k-n+1)(k-n+2)\cdots(k-n+m)\Big]^{1/2}\ ,\\
\nonumber
&&\beta_{r,s}(k)=\Big[(N-k)(N-k-1)\cdots(N-k-s+1)\\
&&\hskip 4cm \times(N-k-s+1)(N-k-s+2)\cdots(N-k-s+r)\Big]^{1/2}\ .
\nonumber
\end{eqnarray}
Since ${\rm Tr}[\rho\, A_1 A_2]$ vanishes by hypothesis, by choosing $n=N-s$, one finally gets the condition:
\begin{equation}
\rho_{nm}\ \Big[m!\, n!\, (N-m)!\, (N-n)!\Big]^{1/2}=\,0\ ,\qquad m<n\ ;
\label{14}
\end{equation}
a similar condition holds for the case $m>n$. As
a consequence, $\rho_{mn}=\,0$ for all $m\neq n$, and thus the original 
density matrix in (\ref{10}) results diagonal
in the Fock representation.%
\footnote{Furthermore, one easily checks that the states (\ref{9}) satisfy the separability criterion for
continuous variable systems introduced in (\cite{Duan}).}

However, most observables of physical interest are non-local with respect to the bipartition
$({\cal A}_1, {\cal A}_2)$ given above. As we shall see in the next Section, the approach to 
phase estimation with BEC-based double-well interferometers is mainly based on 
the following collective bilinear operators:
\begin{equation}
J_x={1\over2}\big(a_1^\dag a_2 + a_1 a_2^\dag\big)\ ,\qquad
J_y={1\over2i}\big(a_1^\dag a_2 - a_1 a_2^\dag\big)\ ,\qquad
J_z={1\over2}\big(a_1^\dag a_1 - a_2^\dag a_2\big)\ ,
\label{15}
\end{equation}
satisfying the $su(2)$ algebraic relations:
\begin{equation}
\big[J_x,\, J_y\big]=i J_z\ ,\qquad
\big[J_y,\, J_z\big]=i J_x\ ,\qquad
\big[J_z,\, J_x\big]=i J_y\ .
\label{16}
\end{equation}
The effect of phase accumulation inside the interferometer can in fact be modeled
through the action of the exponential of a linear combination of these operators. 
It is interesting to notice that,
although the operators in (\ref{15}),
being linear combinations of both $a_1$, $a_1^\dag$ and $a_2$, $a_2^\dag$, are
clearly non-local with respect to the bipartition $({\cal A}_1, {\cal A}_2)$, their
exponentials are not all so. While $e^{i\theta J_x}$ and $e^{i\theta J_y}$, $\theta\in[0, 2\pi]$,
are non-local, the exponential of $J_z$ turns out to be local:
\begin{equation}
e^{i\theta J_z}=e^{i\theta a_1^\dag a_1/2}\cdot e^{-i\theta a_2^\dag a_2/2}\ ,\qquad
e^{i\theta a_1^\dag a_1/2}\in{\cal A}_1\ ,\quad e^{-i\theta a_2^\dag a_2/2}\in{\cal A}_2\ .
\label{17}
\end{equation}
Further, by changing the bipartition of the algebra of all operators $\cal A$,
one can transfer this property to another of the operators in (\ref{15}). For instance,
let us introduce a new set of creation and annihilation operators $b_i^\dag$, $b_i$, $i=1,2$
through the following Bogolubov transformation:
\begin{equation}
b_1={a_1+a_2\over\sqrt{2}}\ ,\qquad
b_2={a_1-a_2\over\sqrt{2}}\ ,
\label{18}
\end{equation}
and their hermitian conjugates. Correspondingly, the three operators
in (\ref{15}) can be equivalently rewritten as:
\begin{equation}
J_x={1\over2}\big(b_1^\dag b_1 - b_2^\dag b_2\big)\ ,\qquad
J_y={1\over2i}\big(b_1 b_2^\dag - b_1^\dag b_2\big)\ ,\qquad
J_z={1\over2}\big(b_1 b_2^\dag + b_1^\dag b_2\big)\ .
\label{19}
\end{equation}
In analogy with what has been done before, one can now define a bipartition 
$({\cal B}_1, {\cal B}_2)$ of the full algebra $\cal A$ using the mode
operators $b_i^\dag$, $b_i$ instead of $a_i^\dag$, $a_i$. In this case, 
it is the exponential of the operator $J_x$ that now turns out to be local:
\begin{equation}
e^{i\theta J_x}=e^{i\theta b_1^\dag b_1}\cdot e^{-i\theta b_2^\dag b_2}\ ,\qquad
e^{i\theta b_1^\dag b_1}\in{\cal B}_1\ ,\quad e^{-i\theta b_2^\dag b_2}\in{\cal B}_2\ .
\label{20}
\end{equation}
This explicitly shows that an operator, local with respect to a given bipartition,
can result non-local if a different algebraic bipartition is chosen.

From the point of view of the states, the above Bogolubov transformation
corresponds to a change of basis in the Hilbert space, from the one
consisting of spatially localized states ($a_i^\dag$ and $a_i$ create and destroy 
particles in the two wells), to the one spanned by their (spatially non-local)
superpositions, as, for instance,
$b_1^\dag |0\rangle=\big[ a_1^\dag|0\rangle + a_2^\dag|0\rangle\big]/\sqrt{2}$ and
$b_2^\dag |0\rangle=\big[ a_1^\dag|0\rangle - a_2^\dag|0\rangle\big]/\sqrt{2}$,
which are energy eigenstates of the Bose-Hubbard Hamiltonian
in the limit of a highly penetrable barrier. 
As a consequence, the Fock states in (\ref{7}) result entangled with respect
to this new bipartition $({\cal B}_1, {\cal B}_2)$; indeed, one finds:
\begin{equation}
|k,N-k\rangle={1\over 2^{N/2}}{1\over\sqrt{k!(N-k)!}}\sum_{r=0}^k\sum_{s=0}^{N-k}
{k\choose r}{N-k\choose s}(-1)^{N-k-s} \big(b_1^\dag\big)^{r+s}\, \big(b_2^\dag\big)^{N-r-s}\, |0\rangle\ ,
\label{21}
\end{equation}
so that $|k,N-k\rangle$ is a combination of $({\cal B}_1, {\cal B}_2)$-separable states. 
A similar conclusion applies to the mixed states (\ref{9}).
This explicitly shows that in the case of systems of identical particles, the notion of
separability can not be given abstractly, but must be associated to a specific bipartition of
the full algebra $\cal A$ of operators.
Because of their physical meaning, in the following
we shall refer to the $({\cal B}_1, {\cal B}_2)$ bipartition as the energy bipartition,
and call spatial bipartition the original $({\cal A}_1, {\cal A}_2)$ one.
We shall now apply these considerations to the problem of phase estimation
in quantum interferometry with systems of identical particles.

\section{Quantum metrology with identical particles}

Ultracold atoms trapped in a double-well optical potential realize
a very accurate interferometric device: indeed, state preparation and beam splitting can be
precisely achieved by tuning the interatomic interaction and by acting on the height
of the potential barrier. The combination of standard Mach-Zhender type interferometric operations,
{\it i.e.} state preparation, beam splitting, phase shift and subsequent beam recombination,
can be effectively described as a suitable ``rotation'' of the initial state $\rho_{\rm in}$
by a unitary transformation \cite{Klauder,Sanders}:
\begin{equation}
\rho_{\rm in}\mapsto \rho_\theta=U_\theta\, \rho_{\rm in}\, U_\theta^\dag\ ,\qquad
U_\theta=e^{i\theta\, J_n}\ .
\label{22}
\end{equation}
The phase change is induced by the generator
\begin{equation}
J_n\equiv n_x\, J_x + n_y\, J_y + n_z\, J_z\ ,\qquad
n_x^2 + n_y^2 + n_z^2=1\ ,
\label{23}
\end{equation}
{\it i.e.} by a general combination of the collective operators introduced in (\ref{15}).
In practice, the state transformation $\rho_{\rm in}\mapsto \rho_\theta$ 
inside the interferometer can be effectively modeled as a pseudo-spin rotation along the
unit vector $\vec{n}=(n_x, n_y, n_z)$, whose choice
depends on the specific realization of the interferometric apparatus
and of the adopted measurement procedure.

General quantum estimation theory allow a precise determination of the
accuracy on which the phase change can be measured in such devices: the accuracy
$\Delta\theta$ with which the phase $\theta$ can be obtained in a measurement involving
the operator $J_n$ and the initial state $\rho_{\rm in}$ is limited by the
following inequality \cite{Helstrom}-\cite{Paris}:
\begin{equation}
\Delta\theta\geq {1\over \sqrt{F[\rho_{\rm in}, J_n]}}\ ,
\label{24}
\end{equation}
where the quantity $F[\rho_{\rm in}, J_n]$ is the so-called quantum Fisher information.
One can show that in general \cite{Luo}
\begin{equation}
F[\rho_{\rm in}, J_n]\leq 4\Big(\Delta_{\rho_{\rm in}} J_n\Big)^2\ ,
\label{25}
\end{equation}
where $\Delta_{\rho_{\rm in}} J_n\equiv\big[ \langle J_n^2\rangle-\langle J_n\rangle^2\big]^{1/2}$ 
is the variance of the operator $J_n$ in the state $\rho_{\rm in}$, the equality holding
only for pure initial states, $\rho_{\rm in}=|\psi\rangle\langle\psi|$.

Given the interferometer, {\it i.e.} given the operator $J_n$ to be measured, one can optimize
the precision with which $\theta$ is determined by choosing an initial state 
that maximizes the corresponding quantum Fisher information.

In the case of distinguishable particles,%
\footnote{Indeed, the inequality involving $F[\rho_{\rm in}, J_n]$ given below
(and similarly the set of inequalities discussed in \cite{Briegel})
has been proved by using the decomposition of the collective
operators $J_\alpha$, $\alpha=x,y,z$, in terms of single-particle pseudo-spin:
$J_\alpha=\sum_i J_\alpha^{(i)}$, where $J_\alpha^{(i)}$ refers to the $i$-th
particle inside the trap. This is possible only if the particles are distinguishable.
In the case of identical particles, single-particle operators like $J_\alpha^{(i)}$
are not addressable, otherwise one would be able to distinguish the particles 
by mean of them: only observables that are symmetric with respect to all permutations
of the $N$ bosons, like the $J_\alpha$'s, are in fact permitted.}
it has been shown that for any separable state $\rho_{\rm sep}$
the quantum Fisher information is bounded by $N$ \cite{Smerzi3}:
\begin{equation}
F\big[\rho_{\rm sep}, J_n\big]\leq N\ .
\label{26}
\end{equation}
This means that by feeding the interferometer with separable initial states, the best 
achievable precision in the determination of the phase shift $\theta$ is bounded
by the so-called shot-noise-limit (also called the standard-quantum-limit):
\begin{equation}
\Delta\theta\geq{1\over\sqrt{N}}\ .
\label{27}
\end{equation}
On the other hand, quite in general, one finds
\begin{equation}
F\big[\rho, J_n\big]\leq N^2\ ,
\label{28}
\end{equation}
so that an accuracy in phase estimation better than the shot-noise-limit 
is in principle allowed, eventually reaching the so-called Heisenberg limit,
$\Delta\theta\geq 1/N$, obtained 
when in (\ref{28}) the equality holds. And indeed, many efforts have been devoted
in order to find
suitable input states $\rho_{\rm in}$ and detection protocols
that would allow such an ultimate sensitivity \cite{Caves1}-\cite{Oberthaler}. 
Notice that, because of (\ref{26}), 
these states must be entangled. Actually, one can turn the argument around and
use the quantum Fisher information for entanglement detection \cite{Smerzi3}; indeed, if for
a state $\rho$ one finds that $F\big[\rho, J_n\big]>N$, than the state is surely entangled.

When dealing with identical particles, as in the case of the condensed bosonic atoms confined
in a double-well optical trap,%
\footnote{When identical bosons are enough far away from each other so that their wavefunction
do not spatially overlap, they effectively behave as distinguishable particles,
and no symmetrization of the total wavefunction is needed \cite{Sakurai}. However, 
this situation is hardly applicable to a gas of condensed ultracold atoms, where all particles
share the same quantum state.}
these conclusions need to be re-qualified.
As discussed in the previous Sections, the notion of separability requires
the choice of an algebraic bipartition; for the case at hand,
the spatial bipartition $({\cal A}_1, {\cal A}_2)$ is a natural one:
local operators are those that can be expressed as the product of 
polynomials in the creation and annihilation operators referring to the two wells.
As shown earlier, with respect to this bipartition, a generic separable mixed
state $\rho$ is diagonal in the Fock basis (\ref{7}), and
can thus be written as in (\ref{9}). For such a state, the quantum Fisher
information can be explicitly computed:
\begin{equation}
F\big[\rho, J_n\big]=
(n_x^2+n_y^2)\bigg[N+2\sum_{k=0}^N p_k\, k(N-k) -4\sum_{k=0}^N {p_k p_{k+1}\over p_k+p_{k+1}}
(k+1)(N-k)\bigg]\ .
\label{29}
\end{equation}
In particular, in the case of a pure state, $\rho_k=|k, N-k\rangle\langle k,N-k|$, this expression
results proportional to the variance of $J_n$ ({\it cf.} (\ref{25})),
\begin{equation}
F\big[\rho_k, J_n\big]=
(n_x^2+n_y^2)\big[N+2 k(N-k)\big]\ ,
\label{30}
\end{equation}
and can always be made greater than $N$ with a suitable choice of $k$.
More specifically, when $\vec n$ lays in the plane orthogonal to the $z$ direction,
so that $n_x^2+n_y^2=1$, one finds that \hbox{$F\big[\rho_k, J_n\big]>N$} for all $k$.
Recalling (\ref{24}), this implies that in this case the phase uncertainty is
smaller than $1/\sqrt{N}$, thus beating the shot-noise-limit.
Actually, when the two wells are filled by the same number of
particles, so that the system is in the state $\rho_{N/2}=|N/2, N/2\rangle\langle N/2,N/2|$,
one can even get close to the Heisenberg limit, since in this case:
\begin{equation}
F\big[\rho_{N/2}, J_n\big]=
{N^2\over2}+N\ .
\label{31}
\end{equation}
Therefore, unlike in the case of distinguishable particles, the quantum Fisher information
can attain a value greater than $N$ even with initial states that are separable with
respect to the spatial bipartition. As a consequence, in general, the
inequality (\ref{26}) does not play any more the role of a separability condition when dealing
with systems of identical particles.

In spite of this, we have just seen that
the accuracy with which the phase change can be determined in interferometers
fed with such separable states can still beat the shot-noise-limit, provided that
the rotation involved in the apparatus is not directed along the $z$ axis.%
\footnote{When ${\vec n}=(0,0,1)$, the quantum Fisher information (\ref{29}) vanishes.
From the physical point of view, this results follows from the fact that the Fock states
in (\ref{7}) are eigenstates of the operator $J_z$, 
so that separability is preserved by rotations generated by it:
$e^{i\theta J_z} |k, N-k\rangle= e^{i\theta(2k-N)} |k, N-k\rangle$.
In other words, in order to take advantage of the improvement in the accuracy 
of the phase determination, one has to use an experimental setup for which
${\vec n}\neq(0,0,1)$.}
As observed before, such rotations are realized by operators that are non-local
with respect to the spatial bipartition.%
\footnote{In fact, only the exponential of $J_z$ happens to be a 
$({\cal A}_1, {\cal A}_2)$-local operator, as shown in (\ref{17}).}
Therefore, given the $({\cal A}_1, {\cal A}_2)$ bipartition, it is not the entanglement
of the states fed into the interferometer that help overcoming the shot-noise-limit
in the phase estimation accuracy; rather, it is the non-local character of the
rotations operated by the apparatus on initially separable states
that allows $\Delta\theta$ to be larger than $1/\sqrt{N}$, with the possibility of closely
approaching the Heisenberg $1/N$ limit.

This result can be physically interpreted in another, equivalent way, which will shed further
light on the notion of separability when dealing with identical particles.
The idea is to change description (and thus bipartition) 
through a suitable Bogolubov transformation, following the discussion
at the end of the previous Section. 

Take the unit vector $\vec n$ to lay
in the plane orthogonal to the $z$ axis, so that one can write
$\vec{n}=(\cos\varphi,\sin\varphi,0)$, $\varphi\in[0,2\pi]$. 
Then, the generator $J_n$ in (\ref{23}) assumes the form:
\begin{equation}
J_n={1\over 2}\Big( e^{-i\varphi}\, a_1^\dag a_2 +e^{i\varphi}\, a_1 a_2^\dag\Big)\ ,
\label{32}
\end{equation}
which clearly shows that its exponential is non-local in the $({\cal A}_1, {\cal A}_2)$ 
bipartition. Nevertheless, it can become local in a different, suitably chosen bipartition.
To this aim, let us introduce a new set of mode operators $b_i^\dag$, $b_i$, $i=1,2$
through the following Bogolubov transformation, that slightly generalizes the one in (\ref{18}):
\begin{equation}
b_1={a_1+e^{-i\varphi}a_2\over\sqrt{2}}\ ,\qquad
b_2={a_1-e^{-i\varphi}a_2\over\sqrt{2}}\ ,
\label{33}
\end{equation}
together with their hermitian conjugates. In this new representation, one has:
\begin{equation}
J_n={1\over 2}\big( b_1^\dag b_1 - b_2^\dag b_2\big)\ ,
\label{34}
\end{equation}
so that the unitary operator that implements the rotation around $\vec n$,
\begin{equation}
e^{i\theta J_n}=e^{i\theta b_1^\dag b_1/2}\  e^{-i\theta b_2^\dag b_2/2}\ ,
\qquad e^{i\theta b_1^\dag b_1/2}\in {\cal B}_1\ ,\quad e^{-i\theta b_2^\dag b_2/2}\in {\cal B}_2\ ,
\label{35}
\end{equation}
is indeed local with respect to the new bipartition $({\cal B}_1, {\cal B}_2)$,
where ${\cal B}_1$ is the subalgebra of polynomials in $b_1^\dag$, $b_1$,
while ${\cal B}_2$ is the one of polynomials in $b_2^\dag$, $b_2$.
In this new language, the state $|N/2,N/2\rangle$
representing the situation of equal filling of the two wells, is
no longer separable with respect to this new bipartition; in fact, one finds ({\it cf.} (\ref{21})):
\begin{equation}
|N/2,N/2\rangle={e^{iN\varphi /2}\over 2^{N/2} (N/2)!}\sum_{k,l=0}^{N/2}
{N/2 \choose k} {N/2 \choose l} (-1)^{N/2-l}\, 
\big({b_1^\dag}\big)^{k+l}\, \big({b_2^\dag}\big)^{N-k-l}|0\rangle\ ,
\label{36}
\end{equation}
while, as seen in the previous Section, any pure $({\cal B}_1, {\cal B}_2)$-separable state must be
a Fock state of the form ${b_1^\dag}^m\, {b_2^\dag}^n |0\rangle$.

Despite these changes, the value of the quantum Fisher information for the initial state
$|N/2,N/2\rangle$ and the observable $J_n$ is unchanged and still given by (\ref{31}),
since it does not depend on the representation used to compute it. This means that
if one is able to build an experimental setup, together with a suitable measure procedure,
which can be modelled in terms of the ``energy'' modes $b_i^\dag$, $b_i$ instead of the original
``spatial'' modes $a_i^\dag$, $a_i$, then the accuracy $\Delta\theta$ with which the phase $\theta$
may be determined can still approach the Heisenberg limit. In such a case, the improvement
in sensitivity with respect to the standard shot-noise-limit is due to the
$({\cal B}_1, {\cal B}_2)$-entanglement of the initial state $|N/2,N/2\rangle$
and not to the non-locality of the transformation that takes place inside the apparatus.

\section{Discussion}

The standard notion of separability for a many-body system made of $N$ distinguishable particles
is based on the natural tensor product structure of the Hilbert space in terms
of the single-particle Hilbert spaces: a state (density matrix) of the $N$-body system 
is separable if it can be written as the convex combinations of products 
of single-particle density matrices.

In the case of a system of identical particles, due to the
symmetrization (or antisymmetrization, for fermions) principle, this definition becomes meaningless.
It can be replaced by a generalized one, that makes use of a ``dual'' language, focusing on the
algebra $\cal A$ of operators of the system instead of the set of its quantum states. One fixes a
partition of $\cal A$ in terms of a set of commuting subalgebras and defines as separable those states
for which the associated expectation values of any factorized 
element of this partition can be written as a convex
combination of products of expectation values.
The notion of separability is thus linked to a specific partition of $\cal A$, so that a given
many-body state can be separable with respect to one partition, but result entangled with respect to a
different one. Nevertheless, this generalized definition of separability reduces
to the familiar one expressed in terms of the single-particle tensor product structure in the case of
a system of distinguishable particles.

We have applied these considerations to the specific case of a system of $N$ ultracold atoms
trapped in an optical double-well potential, whose dynamics is very well captured
by a two-mode Bose-Hubbard Hamiltonian. As we have seen, the second quantized language
makes the application to this case of the new, generalized notion of separability 
very transparent and further allows the discussion of various related issues
in quantum metrology. In fact, through state preparation and trapping potential control, 
this system has been shown to realize a highly sensitive Mach-Zhender interferometer, 
able to measure phase differences with a very high accuracy.

Quite in general, the square error $(\Delta\theta)^2$ in the determination of the phase difference $\theta$
accumulated inside the interferometer is bounded by the inverse of the quantum Fisher information $F$,
whose value can not exceed $N^2$. This gives the smallest possible error in the estimation
of the phase, $\Delta\theta\geq1/N$,
the Heisenberg limit, which, for large $N$, is a huge improvement with respect to 
the standard shot-noise-limit, $\Delta\theta\geq1/\sqrt{N}$.

In the case of a system of distinguishable particles, it has been proven that in order to beat the
shot-noise-limit in the accuracy of the phase determination one needs to feed the
interferometer with suitably $N$-body entangled states. Indeed, 
one can show that for all separable states one has: $F\leq N$; 
as a consequence, the condition $F>N$ signals the presence of entanglement
and at the same time allows $\Delta\theta$ to be smaller than $1/\sqrt{N}$.%
\footnote{Many-body entanglement has also been related 
to spin squeezing \cite{Kitagawa}-\cite{Oberthaler}.
For instance, in the case of distinguishable particles, a state for which the
inequality $N\, (\Delta J_z)^2\geq \langle J_x\rangle^2 + \langle J_y\rangle^2$
is violated is surely entangled and spin squeezed, since the variance of $J_z$
is smaller than the standard quantum limit. Actually, a complete set of inequalities
obeyed by all separable states have been
discussed in \cite{Briegel}: violation of just one of them is enough to signal entanglement.
Let us remark that in the case of a system of identical particles, some of these inequalities
are violated even for separable states, thus loosing their role as entanglement witness.
In this respect, when dealing with systems of identical particles,
spin squeezing does not seem an unambiguous, useful resource for quantum metrology.
A detailed discussion of these points will be reported elsewhere.}

When the interferometer is filled with identical particles, the condition $F>N$ is no longer
a univocal signal of state entanglement. Indeed, we have explicitly seen that in this case
a quantum Fisher information larger than $N$ may be obtained either via a non-local
operation on separable states or via local operations on entangled states. Notice, however,
that the notion of locality {\it vs.} non-locality and that of 
separability {\it vs.} entanglement need always to
be referred to given algebraic bipartitions of the full algebra of observables. 

More specifically, in relation to the actual realization of a BEC-based interferometric experiment
with a double-well optical trap, beam-splitting is usually implemented through
lowering and raising of the inter-well potential barrier, while the subsequent phase estimation
is obtained through number counting of particles inside the two wells. The algebraic
bipartition that is relevant in this case is thus the spatial bipartition
$({\cal A}_1, {\cal A}_2)$, where local observables are those that can be expressed
as the product of operators pertaining to the first and second well, respectively.
A sub-shot-noise accuracy in the determination of the phase difference $\theta$
can then be obtained by acting with a non-local operation, {\it i.e.}
a transformation generated by $J_x$ or $J_y$,
on mode-separable states, {\it e.g.} the balanced Fock state $|N/2,N/2\rangle$.

Nevertheless, as shown at the end of Section 4, a different point of view
can be equivalently adopted: it is based on an alternate measurement protocol,
in which the energy bipartition $({\cal B}_1, {\cal B}_2)$ becomes relevant.
In this case, a local operation suffices 
to get a sub-shot-noise phase estimation accuracy, provided it acts on an entangled initial state.
The practical realization of such a new type of BEC-based interferometer
is surely an interesting experimental challenge.

\vskip 2cm 


\end{document}